\documentclass[journal]{IEEEtran} 
\usepackage{amsmath,amsfonts}
\usepackage{algorithmic}
\usepackage{algorithm}
\usepackage{array}
\usepackage[caption=false,font=normalsize,labelfont=sf,textfont=sf]{subfig}
\usepackage{textcomp}
\usepackage{stfloats}
\usepackage{url}
\usepackage{verbatim}
\usepackage{graphicx}
\usepackage{cite}
\begin{document}

\title{\huge Reconfigurable Airspace: Synergizing Movable Antenna and Intelligent Surface for Low-Altitude ISAC Networks}

\author{Honghao~Wang,
	    Qingqing~Wu,
        Yifan~Jiang,
        Ziyuan~Zheng,
        Ziheng~Zhang,
        Yanze~Zhu,
        Ying~Gao,
        Wen~Chen,
        Guanghai~Liu,
        and Abbas~Jamalipour

\thanks{H. Wang, Q. Wu, Z. Zheng, Z. Zhang, Y. Zhu, Y. Gao, and W. Chen are with the Department of Electronic Engineering, Shanghai Jiao Tong University, Shanghai 200240, China (e-mail: \{hhwang, qingqingwu, zhengziyuan2024, zhangziheng, yanzezhu, yinggao, wenchen\}@sjtu.edu.cn). Y. Jiang is with the State Key Laboratory of Internet of Things for Smart City, University of Macau, Macao 999078, China (e-mail: yc27495@umac.mo), and also with the Department of Electronic Engineering, Shanghai Jiao Tong University, Shanghai 200240, China. G. Liu is with the Research Institute, China United Network Communications Corporation, Beijing 100048, China (e-mail: liugh124@chinaunicom.cn). A. Jamalipour is with the School of Electrical and Computer Engineering, University of Sydney, Australia, and with the Graduate School of Information Sciences, Tohoku University, Japan (e-mail: a.jamalipour@ieee.org). \textit{(Corresponding author: Qingqing Wu.)}}
}

\maketitle

\begin{abstract}
Low-altitude unmanned aerial vehicle (UAV) networks are integral to future 6G integrated sensing and communication (ISAC) systems. However, their deployment is hindered by challenges stemming from high mobility of UAVs, complex propagation environments, and the inherent trade-offs between coexisting sensing and communication functions. This article proposes a novel framework that leverages movable antennas (MAs) and intelligent reflecting surfaces (IRSs) as dual enablers to overcome these limitations. MAs, through active transceiver reconfiguration, and IRSs, via passive channel reconstruction, can work in synergy to significantly enhance system performance. Our analysis first elaborates on the fundamental gains offered by MAs and IRSs, and provides simulation results that validate the immense potential of the MA-IRS-enabled ISAC architecture. Two core UAV deployment scenarios are then investigated: (i) UAVs as ISAC users, where we focus on achieving high-precision tracking and aerial safety, and (ii) UAVs as aerial network nodes, where we address robust design and complex coupled resource optimization. Finally, key technical challenges and research opportunities are identified and analyzed for each scenario, charting a clear course for the future design of advanced low-altitude ISAC networks.
\end{abstract}

\section{Introduction}
\IEEEPARstart{F}{uture} 6G networks are envisaged to seamlessly integrate sensing and communication functionalities to support intelligent services. Unmanned aerial vehicles (UAVs), in particular, are poised to evolve from mere aerial relays into intelligent sensing nodes in the network. Early 5G trials already demonstrated UAVs as new aerial base stations (BSs) or relays to extend coverage, as well as communication users served by terrestrial cells \cite{UAV_Tut_YZeng}. Now, with 6G's emphasis on integrated sensing and communication (ISAC), low-altitude UAV networks are being leveraged for real-time environmental sensing in addition to data relaying \cite{UAV_CL_YFJiang}. The flexible mobility and vantage point of UAVs allow them to rapidly gather data to feed city-scale digital twin models and to enable spatial artificial intelligence (AI) applications that require situational awareness. This burgeoning low-altitude economy, including aerial inspection and autonomous driving coordination, underscores the critical role of UAVs in future transportation infrastructure. ISAC technology has consequently emerged as a key enabler for 6G UAV networks, fusing communication links with sensing frameworks in wireless systems to provide more precise and reliable coverage while mitigating interference among the growing fleets of aerial vehicles.

Two emerging technologies, movable antennas (MAs) and intelligent reflecting surfaces (IRSs), have become dual enablers for reconfiguring the wireless environment in large-scale and small-scale spaces. MA is a mechanically flexible antenna system that actively exploits spatial channel variations via controlled movement or rotation of its elements \cite{ZhuLP_MA_Mag}. By physically moving or rotating antennas (e.g., a UAV-mounted MA array \cite{MA_UAV_XWTang}), an MA-equipped transceiver samples the local electromagnetic field to capture fine-grained channel and echo details that fixed-position antennas (FPAs) would miss. This yields a new degree of freedom (DoF) for spatial diversity: the UAV can seek out antenna positions or orientations that strengthen the desired signal while suppressing interference, without requiring additional radio frequency (RF) chains. In parallel, IRS provides a complementary, passive mechanism to shape the global propagation environment. IRS is typically an electromagnetic reconfigurable surface composed of many subwavelength elements that can electronically adjust the phase and amplitude of reflected radio waves \cite{MA_IRS_Mag}. By intelligently reconfiguring these elements in real time, IRS could create favorable propagation paths on demand, providing more effective channel paths for MAs to achieve constructive signal superposition.

\begin{figure*}[t]
	\centering
	\includegraphics[width=1\textwidth]{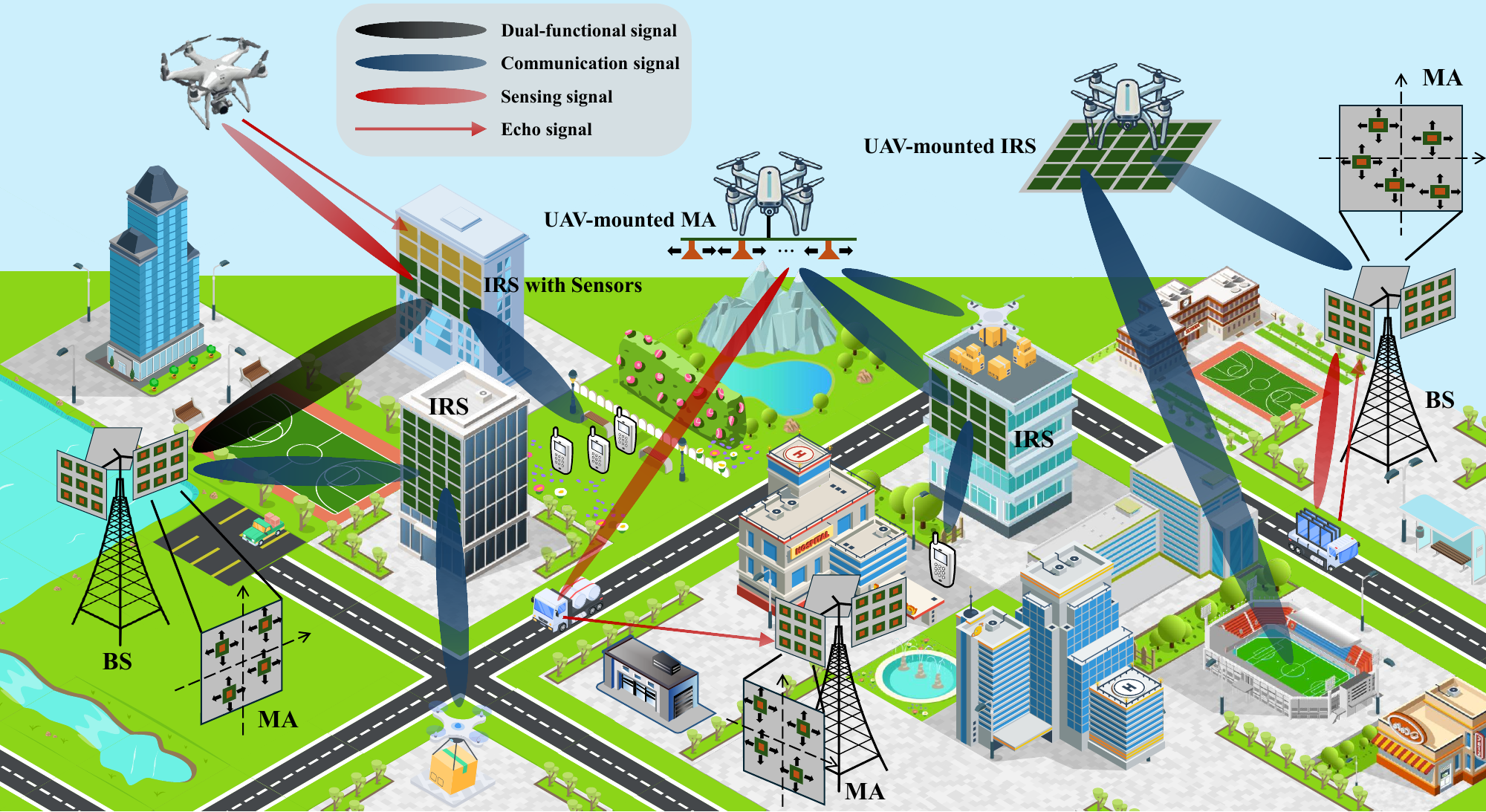}
	\caption{Conceptual vision of MA-IRS empowered low-altitude UAV-ISAC networks.}
	\label{fig1}
\end{figure*}

In concert, the synergy of MA and IRS allows the network to reconfigure both transceiver and environment, significantly boosting dual-function performance. On a large scale, IRSs can construct virtual line-of-sight (LoS) links or focus signal energy over an area. While on a small scale, MAs can fine-tune the beam pattern and antenna directional gain of the transceiver. Recent studies have confirmed that substantial gains can be obtained by deploying MAs, IRSs, or their joint configurations, leading to higher communication throughput \cite{MA_Delay_HHWang}, lower sensing Cramér-Rao bound (CRB) \cite{IRS_Sensing_ZHZHang, MIS_Sensing_ZYZheng}, and improved coverage \cite{MA_IRS_Mag, MA_IRS_YGao}. In this context, a well-designed MA-IRS coordination can simultaneously enhance data transmission and sensing accuracy beyond what either technology could achieve alone. As depicted in Fig. \ref{fig1}, this UAV-ISAC integrated network shows versatile deployments of MAs and IRSs. Specifically, MAs can be equipped on UAVs and ground BSs. By reconfiguring the transceiver beam patterns, they facilitate cooperative sensing of aerial or ground targets while mitigating interference between communication and sensing beams. IRSs can be attached to UAVs or mounted on building facades. Beyond their fundamental function as relays for ensuring communication coverage, they can also serve as new sensing nodes to detect targets. Furthermore, the MA-IRS combination can establish virtual LoS communication links for blocked users at the same time that MAs physically adjust the beam to maximize the radar echo from a target, thereby concurrently improving both the data rate and sensing resolution. Within these scenarios, UAVs exhibit multifaceted roles: they can be communication users, targets to be detected, or act as either active or passive nodes to provide comprehensive communication coverage and sensing services.

The rest of this article is dedicated to exploring how the MA-IRS dual-enabler framework can be harnessed for reconfigurable UAV-ISAC network design. We first elaborate on the design issues and benefits of the network, then present the system architectures and design insights for two core types of UAV deployment scenarios: (i) UAVs as ISAC users, and (ii) UAVs as ISAC network nodes. For each scenario, we discuss the key technical challenges and highlight solution approaches that exploit the MA and IRS reconfigurability to meet the stringent requirements of low-altitude ISAC networks.

\section{MA-IRS-Enabled Reconfigurable UAV-ISAC}
\subsection{Design Issues in UAV-ISAC Networks}
First, the system must fulfill the communication quality of service (QoS) requirements, e.g., ensuring that each user enjoys at least a minimum data rate for reliable service. At the same time, the radar function should detect targets with high probability and estimate their parameters, such as existence, location, and speed, with low error. These dual goals lead to several design objectives and constraints in UAV-ISAC systems.
\subsubsection{Performance metrics} To quantify the performance of ISAC tasks, we first need to determine metrics from both domains. On the communication side, a primary metric is the communication rate, since low-altitude networks often require a minimum rate for each user to guarantee service or maximize the network sum rate. On the sensing side, key metrics include the CRB for estimation accuracy and the detection probability for targets. Typically, the resource allocation, such as power, time, aperture, and ISAC waveform designs, presents a trade-off between communication and sensing \cite{XYJing_UAV_Optimize}.
\subsubsection{Mobility management} UAVs are by nature highly mobile. Although this mobility can be an advantage, since UAVs can be repositioned for better coverage or sensing geometry, it also means that channel conditions vary rapidly with a short coherence time, resulting in the optimal beamforming configuration changing over time. Therefore, the beam directions need to be updated every second or even faster to stay aligned with moving users and targets, where fast channel estimation and beam tracking become critical enablers. The network may employ advanced algorithms, such as extended Kalman filters or machine learning frameworks, to track channel evolution and make predictive adjustments \cite{YPCui_UAV_Track}.
\subsubsection{Interference mitigation} In UAV-ISAC networks, interference occurs in multiple dimensions. Multiple UAVs could interfere with each other, resulting in multiuser communication interference. Radar echoes or high-power probing signals might interfere with communication reception if not properly managed, leading to sensing-communication interference, especially in monostatic ISAC systems. Strategies such as waveform orthogonalization, echo interference cancellation, or time-frequency separation may be necessary to isolate the two functions at receivers. Additionally, when the UAV network coexists with other wireless systems, the network must impose regulatory requirements on beam patterns, transmit power, and spectrum sharing \cite{XWPang_UAV_IRS}.

Thus, UAV-ISAC systems must meet both communication and sensing requirements simultaneously, which requires carefully balancing trade-offs and exploiting every DoF. Integrating MAs and IRS can be an efficient solution to the aforementioned design issues.

\subsection{MA-Enabled Transceiver Reconfigurations}
MA systems utilize the controlled movement or rotation of antennas to harness spatial variations of wireless channels. Rather than treating the propagation environment as fixed and compensating solely through electronic beamforming, MAs sample the electromagnetic field, selecting positions and orientations with favorable link metrics, such as signal-to-noise ratio (SNR) and condition number, without adding RF chains \cite{ZhuLP_MA_Mag}. The concept is naturally generalized to six-dimensional movable antennas (6DMAs), where an element or array is actuated in three-dimensional (3D) position and 3D orientation (pitch, roll, yaw). By reshaping the effective aperture and view angles, 6DMA can exploit amplitude/phase variations, spatial diversity, and, in suitably designed structures, polarization selectivity \cite{6DMA}. Furthermore, the same geometric DoFs that enhance communication can also be co-optimized for sensing: ISAC transceivers can leverage MAs to increase echo signal power and spatial DoFs for data links, thereby obtaining more informative target geometries and reducing communication-sensing coupling, which improves joint performance. Key benefits of MA technology include:
\subsubsection{Array gain} By repositioning and rotating antennas toward the desired signal direction, MAs can align with dominant paths to effectively compensate for path loss. This yields higher receiving power, as the antenna’s radiation pattern is physically steered to capture stronger signal components. Additionally, rotating the transmit antenna surface to face the receiver can significantly enhance the LoS signal and reduce the outage probability.
\subsubsection{Spatial multiplexing} Proper MA positioning can alleviate channel rank deficiencies. In conventional multiple-input-multiple-output (MIMO), closely spaced or fixed antennas often experience correlated channels, which limit the multiplexing gain. In contrast, MAs can be spread out or oriented to view independent scattering clusters, leading to more balanced channel eigenvalues and higher MIMO capacity.
\subsubsection{Interference mitigation} MAs can selectively attenuate interference by maneuvering into positions or orientations that minimize coupling with interfering signals. For instance, an MA-equipped BS can place interferers in nulls of the radiation pattern while pointing toward intended users. This spatial filtering reduces both inter-user interference and potential self-interference between sensing and communication functions.
\subsubsection{Geometric gain} Beyond communication, the geometry between transceivers and targets is crucial to sensing accuracy. Moving antennas to increase the aperture or obtain better angles on targets can achieve a synthetic aperture effect that sharpens the resolution and reduces the estimation error. In other words, the antennas' trajectory can be designed to improve the target observing geometry, yielding a geometric gain in the sensing performance.

\subsection{IRS-Enabled Environment Reconfigurations}
IRS is a metasurface composed of passive elements, each of which can induce a tunable phase shift and amplitude change on incident electromagnetic waves. By coordinating these elements via a software controller, IRS can collectively shape propagation channels, for instance, by beamforming the reflection toward a desired receiver. This enables dynamic reconfigurations of the wireless environment: IRS can redirect signals to overcome blockages, focus energy on specific targets or users, or null out interference, all without requiring additional RF chains.
\subsubsection{From a communication perspective} IRS enables controllable superposition of direct and reflected components. In a UAV-to-ground link with IRS, the user receives two types of signal components: one traveling the direct path from UAV, if not blocked, and one redirected by IRS via a cascaded channel \cite{ZQWei_UAV_IRS}. IRS configuration becomes an explicit design variable: per-element settings of IRS determine how reflected waves are added at receivers, thereby enabling virtual LoS when the direct path is blocked, and enhancing link budget when it is not.
\subsubsection{From a sensing perspective} The round-trip multi-hop sensing channel can be modeled as the target’s radar-cross-section (RCS) and propagation loss on the two-way path. In UAV sensing systems with IRS, the echo strength and detectability depend on the beam alignment between UAV, IRS, and target. When LoS link to the target is blocked, the IRS can serve as a mirror-like relay for the radar probing signal, creating controllable outbound and inbound paths, effectively improving sensing coverage \cite{YXiu_UAV_IRS_ISAC}. If direct LoS path to the target exists, steering IRS main lobe toward the target still increases echo SNR and sharpens angular and range resolution.

Given the above advantages, in MA-IRS systems, the composite channel state, which varies over time, is jointly defined by transceivers’ states, i.e., their antenna positions and orientations, and the environment’s state, i.e., IRS configuration. This raises the issue of a unified spatiotemporal view and the role of MA-IRS co-design. Specifically, wavelength-scale MA repositioning can flip multipath phases from destructive to constructive, while orientation changes reshape antenna patterns and polarization coupling. Element-level phasing on IRS finely controls the relative phase and amplitude of reflected paths, enabling constructive addition at intended points and cancellation toward interferers. This spatiotemporal coupling is particularly the essence of MA-IRS-enabled ISAC. Moreover, by coordinating UAV's pose and IRS deployment, the network actively shapes propagation for both links and sensing geometries, rather than passively accepting random channels.

\begin{figure}[!t]
	\centering
	\includegraphics[width=0.5\textwidth]{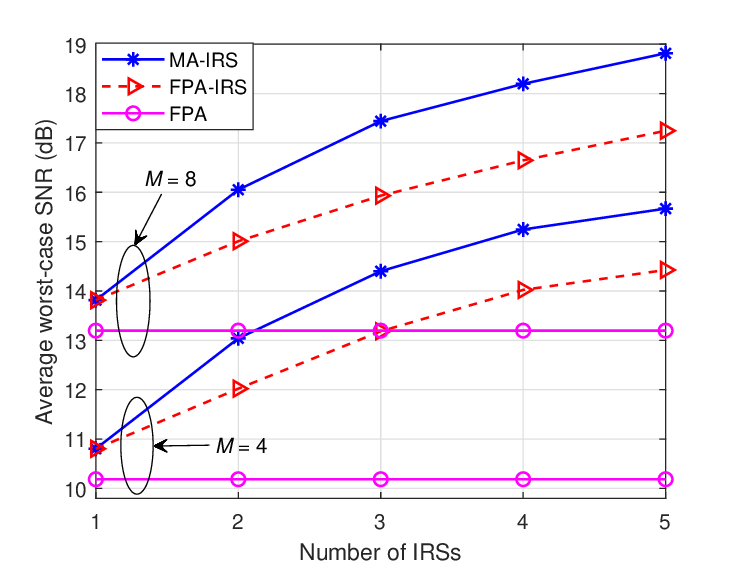}
	\caption{Signal coverage strength enhanced by MA and IRS.}
	\label{Comm}
    \vspace{20pt}
    \includegraphics[width=0.45\textwidth]{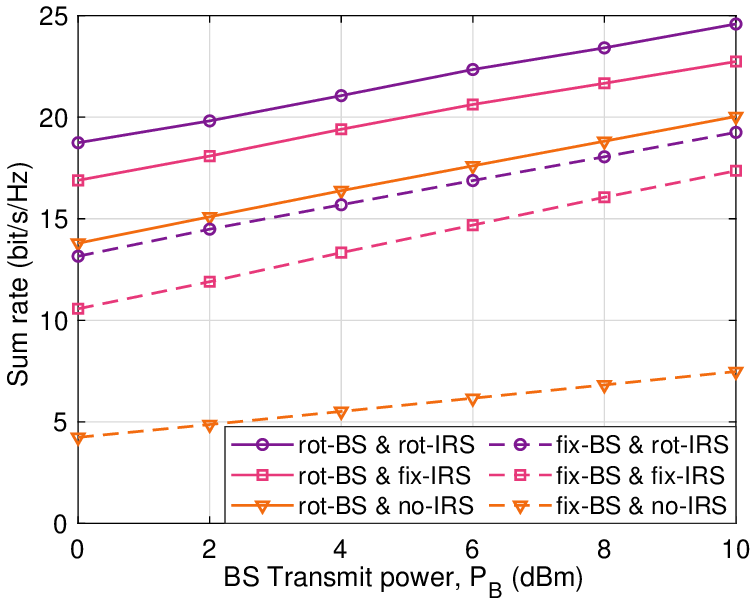}
	\caption{ISAC performance improved by MA and IRS.}
	\label{Rota}
\end{figure}

The quality of signal coverage in a target area is a prerequisite for normal operations of communication or sensing services and stands as one of the most fundamental metrics for evaluating system performance. Fig. \ref{Comm} illustrates via simulation the synergistic effect of MAs at the transmitter and a varying number of IRSs in the environment to enhance signal coverage strength within a specific area. Observations indicate that a configuration with just M=4 MAs combined with 2 IRSs can achieve the same performance as a system with M=8 FPAs (the FPA scheme, represented by pink lines, serves as LoS baselines without any IRS assistance). This demonstrates that within the controllable multipath channels established by multiple IRSs, MA systems can fully leverage multipath phase superposition. This synergy significantly enhances spatial multiplexing, leading to substantial improvements in signal coverage capability while reducing the required number of RF chains/antennas.

Fig. \ref{Rota} considers a 4-rotatable-MA-equipped BS serving 2 users and 2 targets under sensing beam pattern mean-squared-error (MSE) constraints, assisted by IRS. The results show that incorporating either rotatable MAs or IRS individually improves the sum rate compared with the fixed baselines, as each adds spatial flexibility that helps meet sensing constraints with less loss of communication power efficiency. However, the joint MA-IRS rotation configuration achieves a distinctly higher rate, indicating that simultaneously steering MAs and adjusting IRS allows better alignment of both direct and reflected channels while preserving the desired sensing pattern. The cooperative adaptation between MAs and IRS thus enables a more balanced utilization of spatial DoFs, translating into a consistent communication gain under sensing-aware operation.

\section{UAV as ISAC User: High-Precision Tracking and Aerial Safety}
Low-altitude applications impose dual tasks on terrestrial wireless systems: providing reliable, high-throughput communication links and simultaneously performing high-precision tracking to ensure operational safety. However, consumer-grade UAVs typically with low detectability due to their limited sizes and non-metallic components. This chapter investigates novel paradigms that leverage MAs and IRSs to empower the network's ISAC capabilities when UAV is the target of interest.

\subsection{System Models and Configurations}
\begin{figure*}[!t]
	\centering
	\includegraphics[width=1.0\textwidth]{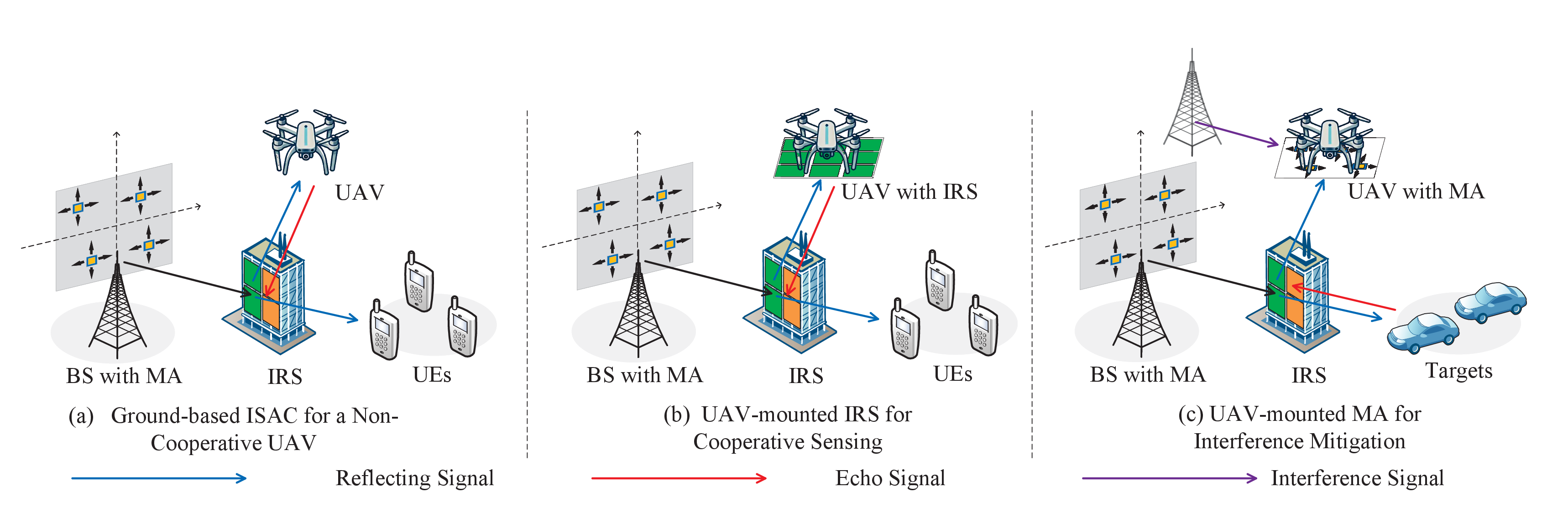}
	\caption{UAV as ISAC user: high-precision tracking and aerial safety.}
	\label{rcc}
\end{figure*}

To comprehensively analyze and fully unlock the potential of reconfigurable technologies for serving aerial users, we first establish a foundational network architecture. In this architecture, the ground infrastructure is always equipped with both MAs at the BS and an IRS in the environment. This powerful ground segment is designed to mitigate unique ground-to-air channel impairments, such as airframe shadowing caused by the UAV's attitude changes. As UAVs typically operate in the air, the channels to ground nodes and other aerial nodes are often dominated by LoS links, which provides a convenience for channel modeling but also raises stricter requirements for precise beam alignment. Building upon this foundational infrastructure, we delineate three distinct operational paradigms, which are differentiated by the capabilities of the UAV itself. These paradigms, illustrated in Fig. \ref{rcc}, range from a baseline non-cooperative scenario to advanced cooperative frameworks, all aimed at achieving high-precision sensing of the targets and maintaining robust communication.

\subsubsection{Ground-based ISAC for a non-cooperative UAV} As depicted in Fig. \ref{rcc}(a), this configuration represents the foundational scenario where the network must sense a standard, non-cooperative UAV that possesses no reconfigurable elements. The ground-based BS, equipped with MAs, performs dual functions of communicating with terrestrial users and sensing the UAV. MAs provide critical spatial DoFs to optimize the channel to the IRS and, in turn, create optimal channel conditions for sensing, effectively enhancing the echo signal quality. Concurrently, the ground-based IRS plays a vital dual role. For communication, it acts as a passive relay, reflecting signals from BS to create virtual LoS links for users, ensuring that the primary aerial sensing task does not degrade ground service. For sensing, it serves two functions: first, it reflects the probing signal toward UAV, and second, it can function as a passive monostatic/bistatic receiver. By receiving the UAV's echo from a different spatial perspective than BS, it provides additional measurements that, when fused with BS's data, can significantly improve localization accuracy and resolve ambiguities. This paradigm enhances network capabilities without requiring any modification to the UAV fleet, making it a highly practical and scalable solution for initial deployment.
\subsubsection{UAV-mounted IRS for cooperative sensing} This paradigm, shown in Fig. \ref{rcc}(b), advances the baseline model by introducing a cooperative onboard IRS, which transforms the UAV from a passive target into an active participant in sensing process. The ground MA-equipped BS transmits probing signals toward the ground IRS, which then reflects the signal to the UAV. Instead of relying on passive scattering, UAV intelligently controls the phase shifts of its mounted IRS to manipulate the reflected echo signal, coherently focusing the energy of probing signal (arriving via ground IRS) into a narrow beam directed back at ground receiver (also potentially via ground IRS). This effectively and artificially boosts the UAV's RCS in the desired direction, leading to a substantial increase in echo SNR. This SNR improvement is paramount for achieving high-precision tracking, as it directly translates to a lower CRB for position and velocity estimates. This model demonstrates a powerful synergy where reconfigurable elements on the target are leveraged to fundamentally enhance sensing performance, particularly beneficial for scenarios requiring centimeter-level accuracy or for tracking at extended ranges where the natural echo would be too weak.
\subsubsection{UAV-mounted MA for interference mitigation} This paradigm, illustrated in Fig. \ref{rcc}(c), represents a fundamental role reversal: UAV is the primary communication user, and the network's sensing function is decoupled and directed towards ground-based targets. The UAV, now acting as a high-priority communication client, is equipped with MAs to enhance its link performance, aiming at proactive interference management in a complex multi-cell environment. By utilizing antennas' movement, UAV can perform real-time spatial filtering at hardware level. This allows it to dynamically adjust its receive radiation pattern, simultaneously creating a high-gain lobe towards the path from its serving BS (via ground IRS) while placing deep nulls in the direction of interfering stations. This user-centric paradigm provides a degree of interference mitigation that is often superior to purely electronic methods, ensuring a robust, high-SNR link. Meanwhile, the sensing task can be performed by independently optimizing the network's resources on the designated ground targets, showcasing a sophisticated decoupling of ISAC functionalities.

\subsection{Core Design Challenges and Opportunities}
The implementation of the aforementioned innovative system models introduces a series of formidable design challenges. Fortunately, the additional spatial DoFs offered by MAs and IRS provide new opportunities to address these issues, but also place higher demands on system design.
\subsubsection{Predictive tracking} High-mobility UAVs operate in rapidly changing wireless channels, where the time it takes to receive feedback for beam adjustment is often longer than the channel remains stable. This issue makes traditional reactive beam tracking ineffective and necessitates a shift to a proactive, predictive framework, whose core is accurately forecasting the UAV's future position and attitude. However, prediction alone is insufficient; the system must act on this forecast. MAs need to be proactively moved/rotated to the predicted optimal positions/orientations, thereby securing the best channel gain ahead of time. In parallel, IRS can precondition the propagation environment by steering reflections towards the UAV's anticipated future location, effectively creating a high-quality virtual LoS link before it is needed. This synergistic combination, the precise predictive framework coupled with proactive optimization by MAs and channel pre-configuration by the IRS, is critical to guarantee the robust connectivity and sensing performance required for mobile UAVs.
\subsubsection{Aerial spoofing} A fundamental security vulnerability in the ISAC paradigm emerges when a low-altitude UAV acts as an undesired intruder. Possessing prior knowledge of the ground-based ISAC station's location, the UAV can be equipped with an IRS to actively suppress its echo signal, thereby evading detection. To counter this, the ISAC station can employ MAs and deploy a friendly IRS in the environment, creating two distinct sensing paths: a direct LoS path and a secondary path via the BS-IRS-UAV link. Specifically, MAs leverage these dual paths for the phase superposition of sensing signals to enhance the strength illuminating the hostile UAV; by forming a bistatic radar system with IRS, it also mitigates the risk of the sensing signal being suppressed along a single direction. Conversely, the IRS carried by the hostile UAV not only suppresses the echo signal strength in the sensing direction but can also capture communication signals from other BSs, redirecting them as interference towards the ISAC station to degrade its sensing SNR. This establishes a direct adversarial relationship between the ISAC station's MA-IRS system and the illegal UAV's onboard IRS, where the relative superiority of this engagement becomes a dynamic struggle, contingent upon the strategic positioning and rotation of MAs, the configuration of ground-based IRS, and the sophisticated design of the illegal UAV's IRS.
\subsubsection{Doppler effect} Distinguishing UAVs from other aerial objects like birds is a critical challenge for airspace security, requiring a shift from basic detection to high-fidelity classification. A key opportunity lies in exploiting micro-Doppler signatures, i.e., unique frequency modulations caused by the rotation of UAVs' propellers. MAs and IRS can significantly enhance the quality of these fine-grained features. An IRS can establish a high-quality virtual LoS link, boosting the target's echo SNR to make the weak micro-Doppler signals more detectable. Meanwhile, the ``field sampling" capability of MAs enables multi-perspective observations of the target, providing richer echo data to construct a more comprehensive feature profile and thereby improve classification accuracy. Conversely, from a communication perspective, the goal is often to eliminate the Doppler effect to ensure link stability. MAs can actively compensate for the UAV's motion by moving in the opposite direction along the LoS. An IRS, on the other hand, can pre-compensate for the Doppler effect electronically. By applying calculated, time-varying phase shifts, IRS can introduce a phase change that is the exact inverse of the Doppler-induced phase shift, ensuring the signal reflected to the receiver is stabilized at its original frequency.

\begin{figure*}[!t]
	\centering
	\includegraphics[width=1.0\textwidth]{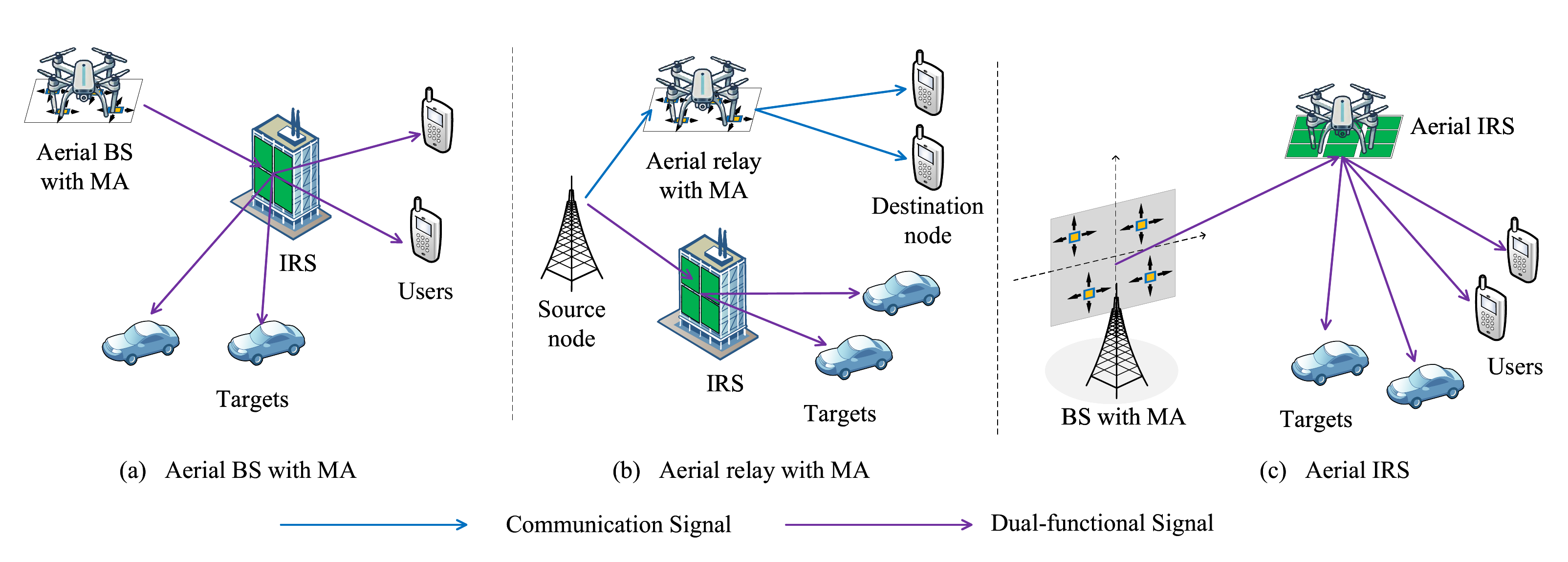}
	\caption{UAV as aerial node: robust design and resource optimization.}
	\label{fig-sec4}
\end{figure*}

\section{UAV as Aerial Node: Robust Design and Resource Optimization}
Besides being a sensing target or communication user, UAVs can also serve as three types of aerial nodes: BS, relay, and IRS carrier, to improve network QoS. In this section, we introduce three system models and key design issues under this scenario configuration, as illustrated in Fig. \ref{fig-sec4}.

\subsection{System Models and Configurations}
\subsubsection{Aerial BS with MA} In traditional UAV networks, UAVs can act as aerial BSs for LoS-dominant channel conditions and flexible wireless coverage, as shown in Fig. \ref{fig-sec4}(a). Nevertheless, the probability for those conditions is usually limited since obstacles could cause blockage and scattering when UAVs are at low altitudes, which degrades the qualities of air-to-ground links. To handle this issue, IRS can be flexibly deployed to create the cascaded BS-IRS-user/target channel, which considerably improves the channel quality. However, fading in the direct BS-user/target channel caused by scattering or multipath propagation still exists. Such unfavorable effects can be suppressed by equipping UAV-mounted MAs, thanks to the antenna position/orientation adjustability. In this way, the direct and cascaded channels can both achieve pure LoS conditions with negligible fading. The combination of IRS and MAs can consequently further improve channel quality compared to merely utilizing IRS to facilitate signal transmission. Moreover, the MA positions/orientations on the aerial BS can be jointly optimized with the reflection coefficients and deployment of IRS, which offers substantial DoFs for system design.
\subsubsection{Aerial relay with MA} When serving as an aerial relay, an MA-equipped UAV can achieve superior communication performance between source and destination nodes to a FPA-equipped UAV, thanks to the improved spatial multiplexing gain. Nevertheless, conventional relaying designs mainly facilitate communication instead of sensing. In other words, a relay UAV with MAs can only improve communication performance while limitedly contributing to sensing in ISAC systems. Besides, enabling UAVs to relay dual-functional signals considerably increases the UAV payload and hardware cost. To counter this, IRS can be dedicatedly utilized to manipulate the sensing/dual-functional signal propagation for sensing performance enhancement, as shown in Fig. \ref{fig-sec4}(b). In addition, the MA positions/orientations of aerial relay and reflection coefficients of IRS can be jointly designed to partially redirect the relaying signals for users/targets so as to strengthen the echo signals while mitigating sensing-communication interference. When the overall ISAC system serves a large number of communication users, IRS can also be exploited to facilitate the communication signal propagation.
\subsubsection{Aerial IRS} In addition to mounting antennas for signal transmitting or relaying, UAVs can carry IRS to facilitate signal propagation by leveraging both their mobility and the LoS-dominant characteristics of air-to-ground channels, as shown in Fig. \ref{fig-sec4}(c). However, the very mobility of aerial IRS introduces rapid channel fluctuations and small-scale fading, which cannot be sufficiently addressed by merely optimizing the reflection coefficients of aerial IRS. A promising solution is to jointly optimize the aerial IRS's reflections with MA positions/orientations of transceivers. This joint approach is achieved by repositioning/rotating MAs to build constructive phase superposition or destructive phase cancellation, actively counteracting the fading caused by IRS's mobility. Furthermore, the optimization needs to be predictive; MA positions/orientations are designed to accommodate both the current and the anticipated future locations of aerial IRS, ensuring robust performance over time. Effective coordination is paramount, requiring precise knowledge of the aerial IRS's trajectory to inform the MAs' proactive adjustments, while the IRS itself can also assist in pre-positioning tasks. This entire process may rely on statistical optimization to manage the high overhead of frequent system reconfigurations.

\subsection{Core Design Challenges and Opportunities}
The MA-IRS framework with UAVs as aerial nodes is anticipated to boost the overall performance of low-altitude ISAC networks. To practically achieve such system performance gain, it is crucial to solve the key design challenges of combining MA, IRS, and UAV, given as follows:
\subsubsection{Unstable UAV hovering/movement} Due to hardware imperfections and environmental variations, such as wind gusts, UAVs cannot maintain stable hovering/movement in the air in practice. As a result, MAs or IRS mounted on UAVs can suffer from channel state information (CSI) errors or jittering, which further incurs inefficient beamforming. In addition, such instability changes rapidly and irregularly, leading to increased difficulty in suppressing such harmful effects. To ensure the signal enhancement resulting from MA/IRS beamforming, it is crucial to design signal processing techniques robust to the above defects. For example, robust MA/IRS beamforming can be designed to guarantee the signal strength over a region of interest in the presence of UAV hovering/movement imperfections. For practical use, the robust MA/IRS beamforming to be designed should have low computational complexity due to the limited computing resources of UAVs.
\subsubsection{Tight design DoFs coupling} The MA-IRS framework, together with UAVs' mobility, enables extremely flexible signal propagation with increased design DoFs, including MA positions/orientations, IRS reflection coefficients, UAV trajectory/pose, etc. The joint optimization of the mentioned spatial or electromagnetic resources promisingly leads to significant performance enhancement. However, it is challenging to obtain a low-complexity solution to this problem due to the tight coupling among these resources. For instance, the UAV pose couples with MA rotations when UAV serves as an aerial BS or relay, resulting in increased computational complexity and deteriorated system performance. To cope with this challenge, the tightly coupled resources need to be decoupled into different spatial and temporal scales, forming easily solvable optimization subproblems at different scales while ensuring the overall network performance. Another promising approach involves leveraging reinforcement learning to optimize the system as a whole; however, this imposes more stringent demands on the generalization capability and robustness of the trained model.
\subsubsection{Clutter and interference} In aerial ISAC systems, echoes received by UAV base stations are typically corrupted by clutter and interference from diverse sources, including other aerial users, inter-cell interference, propeller noise, and self-interference between communication and sensing functions. UAV mobility further complicates mitigation, rendering classical techniques designed for static ground clutter ineffective. This necessitates advanced solutions leveraging MAs and IRS. MAs can dynamically alter their positions to create spatial nulls towards interference sources or induce a controllable Doppler shift to better distinguish the target from slow-moving clutter. Concurrently, IRS reconfigures the wireless environment through passive beamforming, coherently enhancing target signals while nullifying interference. For optimal performance, the position optimization of MAs and IRS beamforming can be jointly designed with techniques such as space-time adaptive filtering and coordinated multi-point transmission.

\section{Conclusion}
This article has advanced a transformative framework integrating MAs and IRS for low-altitude UAV-ISAC networks. The synergy between active MA reconfiguration and passive IRS environment reconstruction effectively countered the core challenges of high mobility and interference, creating vast spatial DoFs to significantly boost both sensing accuracy and communication rates. Our analysis confirmed the framework's versatility, enhancing high-precision tracking when UAVs are ISAC users and enabling robust services when they act as aerial nodes.

\bibliographystyle{IEEEtran}
\bibliography{Ref}

\begin{thebibliography}{10}
\providecommand{\url}[1]{#1}
\csname url@samestyle\endcsname
\providecommand{\newblock}{\relax}
\providecommand{\bibinfo}[2]{#2}
\providecommand{\BIBentrySTDinterwordspacing}{\spaceskip=0pt\relax}
\providecommand{\BIBentryALTinterwordstretchfactor}{4}
\providecommand{\BIBentryALTinterwordspacing}{\spaceskip=\fontdimen2\font plus
\BIBentryALTinterwordstretchfactor\fontdimen3\font minus
  \fontdimen4\font\relax}
\providecommand{\BIBforeignlanguage}[2]{{%
\expandafter\ifx\csname l@#1\endcsname\relax
\typeout{** WARNING: IEEEtran.bst: No hyphenation pattern has been}%
\typeout{** loaded for the language `#1'. Using the pattern for}%
\typeout{** the default language instead.}%
\else
\language=\csname l@#1\endcsname
\fi
#2}}
\providecommand{\BIBdecl}{\relax}
\BIBdecl

\bibitem{UAV_Tut_YZeng}
Y.~Zeng, Q.~Wu, and R.~Zhang, ``Accessing from the sky: A tutorial on {UAV}
  communications for {5G} and beyond,'' \emph{Proc. IEEE}, vol. 107, no.~12,
  pp. 2327--2375, Dec. 2019.

\bibitem{UAV_CL_YFJiang}
Y.~Jiang, Q.~Wu, W.~Chen, and K.~Meng, ``{UAV}-enabled integrated sensing and
  communication: Tracking design and optimization,'' \emph{IEEE Commun. Lett.},
  vol.~28, no.~5, pp. 1024--1028, May 2024.

\bibitem{ZhuLP_MA_Mag}
L.~Zhu, W.~Ma, and R.~Zhang, ``Movable antennas for wireless communication:
  Opportunities and challenges,'' \emph{IEEE Commun. Mag.}, vol.~62, no.~6, pp.
  114--120, Jun. 2024.

\bibitem{MA_UAV_XWTang}
\BIBentryALTinterwordspacing
X.-W. Tang \emph{et~al.}, ``{UAV}-mounted movable antenna: Joint optimization
  of {UAV} placement and antenna configuration,'' \emph{{\rm 2024}, arXiv:
  2409.02469}. [Online]. Available: \url{https://arxiv.org/abs/2409.02469}
\BIBentrySTDinterwordspacing

\bibitem{MA_IRS_Mag}
\BIBentryALTinterwordspacing
Q.~Wu, Z.~Zheng, Y.~Gao, W.~Mei, X.~Wei, W.~Chen, and B.~Ning, ``Integrating
  movable antennas and intelligent reflecting surfaces {(MA-IRS)}:
  Fundamentals, practical solutions, and opportunities,'' \emph{{\rm 2025},
  arXiv: 2506.14636}. [Online]. Available:
  \url{https://arxiv.org/abs/2506.14636}
\BIBentrySTDinterwordspacing

\bibitem{MA_Delay_HHWang}
H.~Wang \emph{et~al.}, ``Throughput maximization for movable antenna systems
  with movement delay consideration,'' \emph{IEEE Trans. Wireless Commun.},
  early access, Jul. 2025, doi: 10.1109/TWC.2025.3587526.

\bibitem{IRS_Sensing_ZHZHang}
Z.~Zhang \emph{et~al.}, ``Multiple intelligent reflecting surfaces
  collaborative wireless localization system,'' \emph{IEEE Trans. Wireless
  Commun.}, vol.~24, no.~1, pp. 134--148, Jan. 2025.

\bibitem{MIS_Sensing_ZYZheng}
\BIBentryALTinterwordspacing
Z.~Zheng, Q.~Wu, Y.~Zhu, W.~Chen, Y.~Gao, and H.~Wang, ``Wireless sensing with
  movable intelligent surface,'' \emph{{\rm 2025}, arXiv: 2509.15627}.
  [Online]. Available: \url{https://arxiv.org/abs/2509.15627}
\BIBentrySTDinterwordspacing

\bibitem{MA_IRS_YGao}
\BIBentryALTinterwordspacing
Y.~Gao \emph{et~al.}, ``Integrating movable antennas and intelligent reflecting
  surfaces for coverage enhancement,'' \emph{{\rm 2025}, arXiv: 2506.21375}.
  [Online]. Available: \url{https://arxiv.org/abs/2506.21375}
\BIBentrySTDinterwordspacing

\bibitem{XYJing_UAV_Optimize}
X.~Jing, F.~Liu, C.~Masouros, and Y.~Zeng, ``{ISAC} from the sky: {UAV}
  trajectory design for joint communication and target localization,''
  \emph{IEEE Trans. Wireless Commun.}, vol.~23, no.~10, pp. 12\,857--12\,872,
  Oct. 2024.

\bibitem{YPCui_UAV_Track}
Y.~Cui \emph{et~al.}, ``Seeing is not always believing: {ISAC}-assisted
  predictive beam tracking in multipath channels,'' \emph{IEEE Wireless Commun.
  Lett.}, vol.~13, no.~1, pp. 14--18, Jan. 2024.

\bibitem{XWPang_UAV_IRS}
X.~Pang, W.~Mei, N.~Zhao, and R.~Zhang, ``Intelligent reflecting surface
  assisted interference mitigation for cellular-connected {UAV},'' \emph{IEEE
  Wireless Commun. Lett.}, vol.~11, no.~8, pp. 1708--1712, Aug. 2022.

\bibitem{6DMA}
X.~Shao and R.~Zhang, ``{6DMA} enhanced wireless network with flexible antenna
  position and rotation: Opportunities and challenges,'' \emph{IEEE Commun.
  Mag.}, vol.~63, no.~4, pp. 121--128, Apr. 2025.

\bibitem{ZQWei_UAV_IRS}
Z.~Wei \emph{et~al.}, ``Sum-rate maximization for {IRS}-assisted {UAV} {OFDMA}
  communication systems,'' \emph{IEEE Trans. Wireless Commun.}, vol.~20, no.~4,
  pp. 2530--2550, Apr. 2021.

\bibitem{YXiu_UAV_IRS_ISAC}
Y.~Xiu \emph{et~al.}, ``Secure enhancement for {RIS}-aided {UAV} with {ISAC}:
  Robust design and resource allocation,'' \emph{IEEE Trans. Veh. Technol.},
  pp. 1--16, early access, Oct. 2025, doi: 10.1109/TVT.2025.3623453.

\end{thebibliography}

\vfill
\end{document}